\documentclass[10pt,letterpaper]{article}
\usepackage[T1]{fontenc}
\usepackage[utf8]{inputenc}
\usepackage[margin=0.85in]{geometry}
\usepackage{mathptmx}
\usepackage{graphicx,longtable,booktabs,array,caption,xurl}
\usepackage[hidelinks]{hyperref}
\usepackage{fancyhdr}
\hypersetup{pdftitle={After the Award: The Authorization Gap in Academic Access to Frontier AI},pdfauthor={Olga Lavinda}}
\begin{document}
{\LARGE\bfseries After the Award: The Authorization Gap in Academic Access to Frontier AI\par}
\vspace{10pt}
{\large Olga Lavinda\par}
\vspace{3pt}
Department of Chemistry and Biochemistry, Stern College for Women, Yeshiva University\par
New York, NY 10016, USA\par
Correspondence: \href{mailto:olga.lavinda@yu.edu}{olga.lavinda@yu.edu}\par
\href{https://orcid.org/0000-0003-3577-5984}{ORCID: 0000-0003-3577-5984}
\vspace{5pt}
\section*{Abstract}
Academic access programs distribute frontier artificial intelligence as research infrastructure, but awards and institutional authorization are distinct stages. We examine the handoff through structured coding of 15 publicly documented access pathways captured on 19 September 2026, of which ten met prespecified inclusion criteria, and a process trace at one US university. Two isolated model-assisted coding passes found explicit eligibility criteria in nine programs and three implied institutional prerequisites across the corpus. Downstream specification was weaker: access duration was unstated in six programs, seven reported no use or outcome metric, and liability assignment was unstated in five to seven. Public terms did not establish an unambiguous institution-independent path from award to intended use in eight to nine programs, a composite that includes possible and unclear cases. In the institutional trace, policy required review even for free tools. A bounded request generated a ticket but no substantive response or decision pathway during the observation window. We define the authorization gap as the distance between an access award and authorized research use. Tracking first use, clearance, time to first use, and persistence would help programs distinguish allocated resources from usable scientific infrastructure.

\textbf{Keywords:} frontier AI access; research infrastructure; institutional governance; computing equity; science policy; higher education

\section*{1. Introduction}
Frontier artificial intelligence is increasingly distributed to scientists as research infrastructure. Model developers, universities, consortia, and public agencies now offer free seats, API credits, cloud credits, verification routes, and compute allocations to academic and nonprofit researchers. Anthropic, for example, announced 10,000 free or discounted seats for verified scientific research groups, while separate programs provide project credits and specialized access for eligible life-science organizations [1,2]. OpenAI, academic institutions, and public programs offer related routes through researcher credits, workspace access, or compute allocation [3-10].

These programs are commonly described and counted at the point of award. Seats are opened, credits are allocated, or projects are approved. Yet scientific use occurs inside institutions with their own rules for procurement, information security, student participation, research data, contracts, and accountability. An award to a principal investigator may therefore be usable immediately, usable only after an institutional action, or left in an unresolved state that neither the provider nor the institution measures.

This paper asks: \textbf{What do public academic AI-access programs specify about the transition from an award to authorized research use?} The question is narrower than whether universities have AI policies and different from whether a model is technically available. It concerns the administrative joint between the two.

We define the \emph{authorization gap} as the distance between allocation of AI access to a researcher and authorization of the intended research workflow within the receiving institution. The gap may contain time, additional approvals, unassigned responsibility, or the absence of an identifiable decision pathway. It is a joint-system property: the provider controls the award and account mechanism, while the institution controls which uses are authorized under its policies. Neither side can observe the full process from award counts alone.

\begin{figure}[!htbp]\centering
\includegraphics[width=\linewidth]{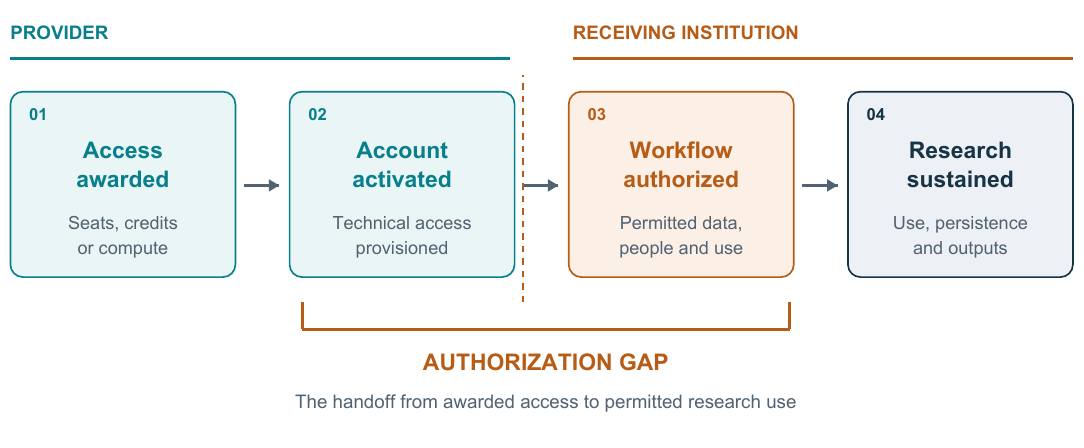}
\caption{From award to authorized research use. The schematic separates provider-level allocation and account activation from institutional clearance and sustained research use. The authorization gap is the unresolved handoff between these systems; the order of activation and clearance may vary. The diagram is conceptual and does not encode measured transition rates.}
\end{figure}
Our initial hypothesis was that access programs would contain numerous hidden institutional prerequisites. The evidence did not support that hypothesis. Both coding passes found only three such prerequisites under a deliberately strict rule. Instead, the corpus showed an asymmetry: admission was usually specified, while activation, accountability, duration, and measures of successful use were frequently silent or indeterminate. A process trace then showed how that silence can meet an institutional policy that requires review but does not expose an operational path to a decision.

The paper makes three contributions. First, it treats academic AI-access program documents as governance artifacts and provides a reproducible coding frame. Second, it distinguishes allocation from authorization and identifies the handoff as a measurable site of unequal research capacity. Third, it proposes a minimal set of program-level indicators that would reveal whether awarded access becomes usable scientific infrastructure.

\section*{2. Related work}
\subsection*{2.1 Governance artifacts and responsibility}
Terms of service, program notices, and application rules do more than describe a product: they allocate responsibility and shape what users can do. Pandit and colleagues coded the terms of six consumer generative-AI services and identified substantial responsibility placed on users despite limited user control [11]. Treude similarly analyzed terms across AI coding tools, finding a recurrent transfer of responsibility for correctness, safety, and legal compliance to developers [12]. These studies establish public program documents as empirical objects through which accountability can be examined.

Grant documents provide a related precedent. Bateyko and Levy analyzed more than 40,000 federal notices of funding opportunity and manually reviewed 407 notices that mentioned AI [13]. They found that agencies promoted AI while rarely using grant conditions to govern its downstream use. Their result directs attention to a general asymmetry between allocating a resource and specifying conditions for what follows.

The present study extends governance-artifact analysis to academic frontier-AI access programs. Its distinct object is the handoff between the allocating program and the receiving research institution.

\subsection*{2.2 Higher-education AI governance}
University AI-policy research has concentrated largely on teaching, learning, academic integrity, and data protection. Wang and colleagues analyzed policies and resources from 100 US universities and found broad institutional activity but strong emphasis on educational use, privacy, and academic integrity [14]. More recent work across institutions in 34 states reports divergence between university-level policies emphasizing security and risk and school-level policies emphasizing pedagogy and tool use [15]. The resulting fragmentation is especially consequential for research workflows, which cross instructional, information-technology, procurement, and research-administration functions.

Professional guidance increasingly recommends institution-level approval of specific AI tools so that users can identify compliant systems and avoid duplicative review [16]. Survey and regulatory work also reports a gap between institutional policy development and researchers' awareness of applicable rules [17]. These studies characterize institutional governance maturity. They do not examine whether external access programs document the point where their awards enter those institutional systems.

\subsection*{2.3 Access and the unit of success}
Public and private initiatives seek to broaden access to advanced models and compute. Their natural administrative units are awards, seats, credits, and allocations. Those units are necessary but incomplete measures of scientific reach. A seat that cannot be cleared for the intended group, a credit award that cannot be activated, or an account that cannot be used with the relevant data may be counted as access even when no research workflow begins.

The authorization gap therefore concerns measurement as much as approval. It asks whether the success metric follows the resource beyond allocation to activation, sustained use, and research output.

\section*{3. Methods}
\subsection*{3.1 Study 1: program corpus}
We constructed a candidate corpus of 15 academic and nonprofit research access pathways. A pathway was included when, on the capture date, it (1) provided frontier models, model APIs, or explicitly AI-capable compute; (2) named academic or nonprofit researchers as eligible or as the target population; (3) exposed public terms without a login wall, nondisclosure agreement, or mandatory sales contact; and (4) accepted applications or stated a future or recurring cycle.

All sources were captured on 19 September 2026. Ten pathways met the criteria: six operated by frontier-model laboratories, one public-agency compute allocation, and three institution- or consortium-mediated pathways. Two generic cloud-credit programs were retained as context exclusions because their captured pages did not explicitly name AI/ML, frontier models, or AI-capable compute. Three additional programs were excluded because their application deadlines had passed before the capture date.

The unit of analysis was a publicly documented access pathway. Distinct routes from one provider were separated when recipient, application route, capability, or verification rule differed materially.

\subsection*{3.2 Coding frame and procedure}
A ten-dimension codebook recorded: unit of grant (D1), stated eligibility (D2), institution-type filter (D3), implied institutional prerequisites (D4), liability assignment (D5), documentation burden (D6), allocation-to-use mismatch (D7), term duration (D8), reported success metric (D9), and public discoverability (D10). Absence of information was coded as an observation rather than left missing.

D4 used a strict rule: an implied prerequisite required a specific action or capacity demanded by the public procedure, not stated as an eligibility criterion, and not completable by the researcher alone. D7 asked whether the named awardee could use the access for the intended workflow without additional institutional action. Ambiguous or partial activation language was coded \texttt{possible} or \texttt{unclear}, not converted into a confirmed barrier.

The frame was piloted on three programs spanning the represented sectors, revised once, and frozen before full coding. Two isolated model-assisted coding passes then evaluated all ten included programs from the frozen PDF captures. Each pass had access only to the codebook, manifest, blank coding sheet, and source PDFs; neither received the other pass or preliminary notes. Agreement was calculated after both sheets were complete. Because some interpretive dimensions showed modest agreement, the Results report exact agreement and coder-specific ranges rather than forcing post hoc consensus. Direct quotations supporting D3, D4, D5, and D7 are retained in the supplementary coding sheets.

Analyses were descriptive and prespecified. One proposed analysis subtracting the number of implied prerequisites from stated eligibility conditions was dropped before manuscript reporting because the counts describe different objects. A proposed capability-tier analysis was not performed because the frozen data did not include a prespecified tier variable.

\subsection*{3.3 Study 2: institutional process trace}
Study 2 traced one awarded scientific-team workspace at a US university. The site is not named in the case description, and operational identifiers are excluded. The author was the principal investigator and workspace administrator.

Evidence comprised four artifact classes: the provider's public program terms; the receiving institution's faculty/staff AI policy; the researcher's submitted request; and the automated incident receipt. Names, contact information, ticket identifiers, and message-reference strings were excluded from the analytical record. The trace reports dated system states and document contents, not the motives or conduct of identifiable employees.

The institutional policy instructed faculty and staff to contact information technology services before purchasing or acquiring even free AI products and stated that vendor management would route the request for validation. It also prohibited use of internal, restricted, or personal information without express permission. The researcher's request therefore specified supervised use with scientific literature, public datasets, computational workflows, and other public or non-restricted research information, explicitly excluding confidential or restricted institutional research data. The request asked whether the use was consistent with policy and whether an approval process existed.

The request date was supplied by the author and is not visible in the retained screenshots; interval calculations using that date are labeled accordingly. The observation window lasted 18 days. No human participants were recruited, and no identifiable communication content was analyzed.

\section*{4. Results}
\subsection*{4.1 Corpus and stated admission rules}
Ten of 15 candidate pathways met the inclusion rule (Table 1). All ten included programs were publicly discoverable through an official page or stable official program URL. Both coding passes found that 9 of 10 stated at least one explicit eligibility condition. They agreed exactly that 5 of 10 imposed a hard institution-type filter and 5 imposed none.

The original hidden-prerequisite hypothesis was not supported. Both coders identified three implied institutional prerequisites across the full corpus. They differed on which non-pilot program supplied one of those instances, but not on the total. By contrast, their granular counts of stated eligibility conditions were 54 and 37. Agreement on the exact D2 count was low (Cohen's $\kappa$ = 0.231), largely because multi-part application forms could be counted as one required artifact or decomposed into distinct positive conditions. The direction of the result did not depend on that choice: public terms contained many stated admission conditions and few strictly inferable institutional prerequisites.

\par\noindent\begin{minipage}{\linewidth}
\textbf{Table 1. Included program pathways.}
\par\vspace{5pt}
{\footnotesize\setlength{\tabcolsep}{4pt}\renewcommand{\arraystretch}{1.18}
\begin{tabular}{>{\raggedright\arraybackslash}p{\dimexpr 0.06462\linewidth-2\tabcolsep\relax}>{\raggedright\arraybackslash}p{\dimexpr 0.46154\linewidth-2\tabcolsep\relax}>{\raggedright\arraybackslash}p{\dimexpr 0.23538\linewidth-2\tabcolsep\relax}>{\raggedright\arraybackslash}p{\dimexpr 0.23846\linewidth-2\tabcolsep\relax}}
\toprule
\textbf{ID} & \textbf{Program pathway} & \textbf{Operator sector} & \textbf{Unit} \\
\midrule
P01 & OpenAI Researcher Access Program & Frontier lab & Credits \\
P02 & ChatGPT for Academic Researchers & Frontier lab & Seats \\
P03 & Anthropic External Researcher Access Program & Frontier lab & Credits \\
P04 & Claude Team Plan for Scientists & Frontier lab & Seats \\
P05 & Anthropic AI for Science project credits & Frontier lab & Credits \\
P06 & Anthropic Life Sciences Verification Program & Frontier lab & Mixed \\
P09 & NAIRR Pilot research resource allocations & Public agency & Compute allocation \\
P10 & University of Toronto DSI Claude API Credit Award & Academic institution & Credits \\
P14 & Connecticut AI Alliance AI Computing Access Program & Academic consortium & Credits \\
P15 & Stanford HAI Google Cloud Credit Grants & Academic institution & Credits \\
\bottomrule
\end{tabular}
}\end{minipage}\par\vspace{6pt}
\subsection*{4.2 The downstream specification gap}
The stronger pattern appeared after admission and allocation (Table 2; Figure 2). Duration was unstated in 6 of 10 programs in both coding passes. Seven of ten reported no use or outcome metric; this count treats reported numbers of awards or organizations onboarded as allocation metrics, not evidence of use or scientific outcome. Only three programs requested or reported use, outputs, or multiple metric families.

Liability or accountability for use and added participants was not stated in 5 of 10 programs in one coding pass and 7 of 10 in the other. The difference arose from whether account-administration duties were interpreted as assignment of liability. Public terms did not establish an unambiguous, institution-independent path from award to intended use in 8 of 10 programs in one pass and 9 of 10 in the other. This range includes confirmed, possible, and unclear mismatches; it does not assert that all of those pathways encounter an institutional barrier in practice.

\par\noindent\begin{minipage}{\linewidth}
\textbf{Table 2. Cross-coder results for core dimensions.}
\par\vspace{5pt}
{\footnotesize\setlength{\tabcolsep}{4pt}\renewcommand{\arraystretch}{1.18}
\begin{tabular}{>{\raggedright\arraybackslash}p{\dimexpr 0.42308\linewidth-2\tabcolsep\relax}>{\raggedright\arraybackslash}p{\dimexpr 0.10769\linewidth-2\tabcolsep\relax}>{\raggedright\arraybackslash}p{\dimexpr 0.10769\linewidth-2\tabcolsep\relax}>{\raggedright\arraybackslash}p{\dimexpr 0.36154\linewidth-2\tabcolsep\relax}}
\toprule
\textbf{Indicator} & \textbf{Coding pass 1} & \textbf{Coding pass 2} & \textbf{Interpretation} \\
\midrule
Programs with explicit eligibility criteria & 9/10 & 9/10 & Admission usually specified \\
Hard institution-type filter & 5/10 & 5/10 & Exact agreement \\
Total implied prerequisites & 3 & 3 & Original prevalence hypothesis not supported \\
Liability not stated & 5/10 & 7/10 & Account-administration language drove disagreement \\
Duration not stated & 6/10 & 6/10 & Exact substantive result \\
No use or outcome metric & 7/10 & 7/10 & Allocation counts excluded as outcomes \\
Direct path from award to use not established & 8/10 & 9/10 & Possible/unclear/confirmed mismatch combined \\
Publicly discoverable & 10/10 & 10/10 & Exact agreement; $\kappa$ undefined because no variance \\
\bottomrule
\end{tabular}
}\end{minipage}\par\vspace{6pt}
\begin{figure}[!htbp]\centering
\includegraphics[width=\linewidth]{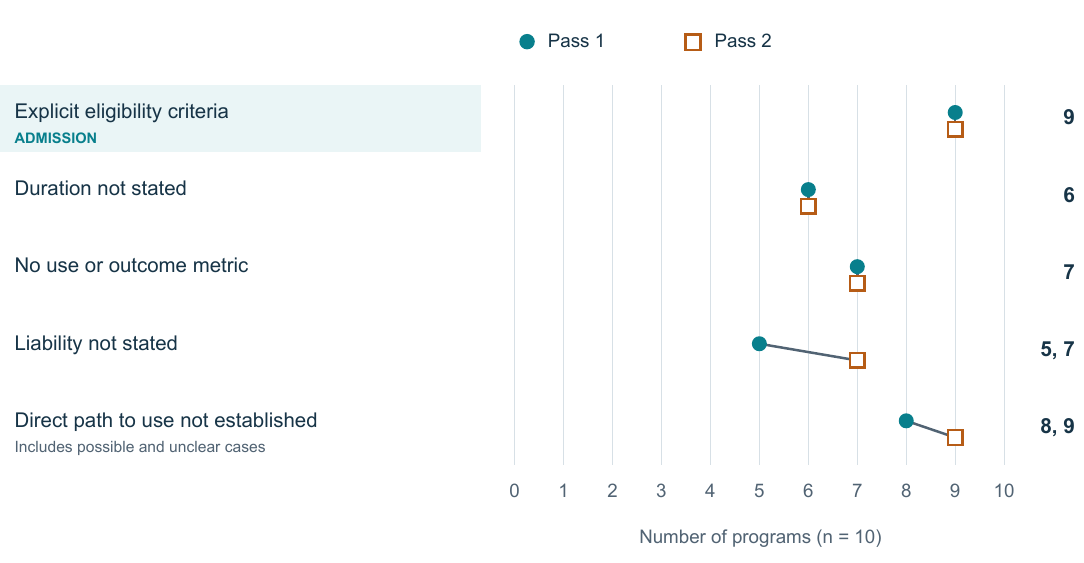}
\caption{Public documentation across ten included programs. Circles and squares show coding passes 1 and 2; each point is a program count on a common zero-to-ten scale. Connecting lines mark coder differences, not confidence intervals. The eligibility row describes admission, while the remaining rows describe downstream omissions or ambiguity. The final row combines possible, unclear, and confirmed allocation-to-use mismatch; it is not a count of demonstrated barriers.}
\end{figure}
Agreement was perfect for D1, D3, and normalized D8 ($\kappa$ = 1.000). It was moderate for D6 ($\kappa$ = 0.559) and D9 ($\kappa$ = 0.492), lower for D4 ($\kappa$ = 0.412) and D5 ($\kappa$ = 0.355), and lowest for D7 ($\kappa$ = 0.275) and the granular D2 count ($\kappa$ = 0.231). The paper therefore treats direct, high-agreement dimensions as point estimates and preserves ranges for the interpretive dimensions.

\subsection*{4.3 Study 2: policy without an operational decision path}
The provider's public terms stated that, after verification, a principal investigator could add researchers in the lab to the scientific team plan [2]. The receiving institution's existing AI policy created a separate obligation: faculty and staff were instructed to contact information technology services before acquiring even free AI products. The policy stated that vendor management would route the request for validation.

At the beginning of the observation window (Day 0), the author reports submitting a request through the designated support channel. The request described the awarded 25-seat scientific workspace, the intended participation of supervised student researchers, and a bounded workflow limited to public or non-restricted scientific information. It asked whether that use complied with institutional policy and what review or approval would be required. The system created an incident ticket.

By the evidence cutoff 18 days later, no substantive response, documentation request, authorization decision, responsible decision function, or alternative pathway had been provided. The observed interval was therefore right-censored at Day 18, calculated from the author-reported request date. No subsequent substantive administrative action was observable in the retained evidence beyond automated ticket creation.

\begin{figure}[!htbp]\centering
\includegraphics[width=\linewidth]{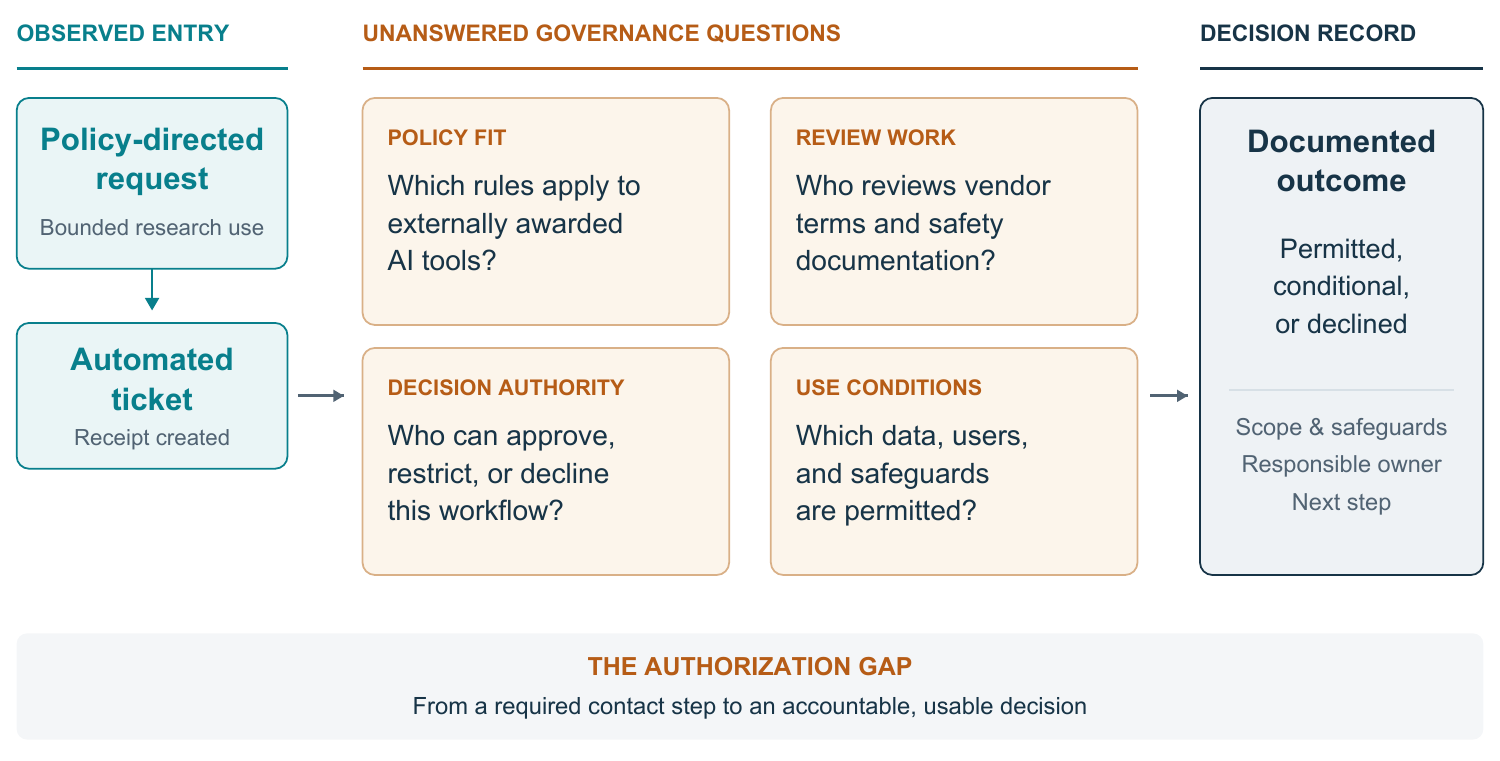}
\caption{The authorization gap as an unresolved governance handoff. The observed policy-directed request produced an automated ticket. The central questions identify the policy interpretation, documentation review, decision authority, and use conditions needed to connect that entry point to a documented decision. The center and right panels are an analytical framework; the retained case evidence establishes the entry point and unresolved authorization state, without measuring internal review activity or workload.}
\end{figure}
The case does not show that the institution rejected the access, that an answer did not exist internally, or that any individual acted improperly. It shows that a researcher following the published policy could reach the specified entry point without reaching a documented authorization state. The provider could count the workspace as awarded while the institution could count the request as received, even though the intended team workflow remained unresolved.

\section*{5. Discussion}
\subsection*{5.1 The authorization gap is downstream of eligibility}
The initial hypothesis focused on hidden prerequisites in program admission. Its lack of support redirected attention to the downstream handoff. Academic AI-access programs were generally explicit about who could apply, and their terms were discoverable. What was frequently absent was downstream specification: who authorizes group use, how long access lasts, who is accountable for added participants, and what counts as successful use.

This distinction matters for equity. Program selection occurs before allocation; institutional capacity acts after it. Institutions with established enterprise agreements, security review, procurement, and named AI-governance functions may convert an award into use routinely. At institutions without a documented route, the first researcher must discover or create the pathway while the award term is already running. Equal awards can therefore produce unequal usable access.

\subsection*{5.2 Policy presence is not pathway availability}
Study 2 is not a case of total policy absence. The institution had a detailed AI policy and the researcher followed it. The unresolved point was operational: the policy assigned a duty to contact information technology services and promised routing, but the reviewed documentation did not expose decision criteria, a responsible endpoint, an expected service interval, or an escalation route for externally awarded research access.

Making that pathway operational may require substantial human judgment. An institution may need to determine how an externally awarded tool fits existing policy, review vendor terms and safety documentation, and decide which research or educational uses, participants, and data are permitted under appropriate safeguards (Figure 3). These are plausible review demands rather than measured causes of the unresolved request. The governance question is how those tasks acquire an accountable owner and produce a decision that the researcher can act on.

That distinction advances current discussion of university AI governance. Policy inventories can determine whether a document exists and what risks it names. They do not necessarily reveal whether a faculty researcher can obtain a decision for a concrete workflow. Authorization-path availability should be measured separately from policy presence.

Public community discussions are consistent with demand for clearer institutional guidance, but they are not prevalence evidence. Recent faculty threads ask whether universities have rules for professors' own AI use, while education-community discussions describe institutions operating with fragmented or unpublished guidance [18,19]. A useful contrast is the University of Maryland's public documentation of individual, purchased, and managed Claude pathways, including where central support is not available [20,21]. These examples show that the pathway can be documented without requiring a uniform decision across institutions.

\subsection*{5.3 What access programs should measure}
Five additions would make the allocation-to-authorization handoff observable:

\begin{enumerate}
\item \textbf{First-use rate:} the proportion of awards that reach first substantive use.
\item \textbf{Time to first use:} median and distribution from award to first substantive use, reported separately from application-review time.
\item \textbf{Institutional-clearance state:} whether the recipient reports that the intended workflow is institutionally cleared, pending, not required, or unresolved.
\item \textbf{Persistence:} whether the proposed workflow remains active at a fixed interval such as 90 days.
\item \textbf{Accountability disclosure:} the party responsible for added participants, data restrictions, and workspace administration.
\end{enumerate}
These fields do not require programs to decide institutional policy. They allow programs to distinguish nominal distribution from usable access and to identify where support or documentation is needed.

Receiving institutions can make the same handoff tractable by publishing a functional route for externally awarded AI access: the entry point, decision owner, required evidence, data categories permitted, expected response interval, and appeal or escalation path. The aim is not automatic approval. It is a decision pathway proportionate to the scientific workflow and data involved.

\section*{6. Limitations}
The corpus is a dated snapshot of a rapidly changing program landscape and contains ten included pathways, predominantly from the United States. Public terms may differ from negotiated or operational practice. The coding frame is strongest for directly observable dimensions; lower agreement for D2, D5, and D7 shows that responsibility and allocation-to-use language remain interpretive even under a frozen codebook. We preserve coder-specific ranges and supporting quotations for that reason.

Study 2 is one purposefully selected case and demonstrates a mechanism, not its frequency. The author was a participant in the process. Reporting is therefore limited to dated artifacts and observable system states. To reduce identification risk, the unredacted policy and ticket artifacts are withheld; the public package supplies deidentified evidence descriptions and the coding logic. The 18-day observation window, calculated from the author-reported request date, is right-censored at the 27 September 2026 evidence cutoff.

Community posts and institutional comparison pages were used only as contextual evidence. They were not sampled systematically and do not support prevalence claims.

\section*{7. Conclusion}
Academic access programs are widening the nominal availability of frontier AI for science. Their public rules are comparatively strong at describing who may enter. They are weaker at showing whether an award becomes institutionally authorized, accountable, sustained research use.

The authorization gap names this missing handoff. In the program corpus, admission criteria were common and hidden prerequisites were sparse, while duration, accountability, allocation-to-use pathways, and outcome measurement were frequently absent or unresolved. In the institutional trace, a detailed AI policy generated a mandatory contact step but not an observable decision pathway within the study window.

Programs that count awards without first use and clearance cannot distinguish distributed access from usable infrastructure. Measuring the handoff would make visible where scientific opportunity may be delayed or lost after allocation - and where a small amount of operational documentation could recover it.

\section*{Ethics, competing interests, and disclosure}
The study analyzed public program materials and administrative artifacts. No human participants were recruited and no identifiable individual-level content was analyzed. The author holds access through one program in the corpus. The provider had no role in study design, analysis, writing, or the decision to disseminate this work. The author declares no competing financial interest.

Two isolated model-assisted passes coded the frozen public corpus. The passes were not exposed to one another's outputs. All reported program evidence is traceable to captured source text; disagreements are reported rather than hidden through unreported consensus. Generative AI assisted manuscript organization and editing under author direction. The author is responsible for the final text, citations, and interpretation.

\section*{Funding}
No external funding.

\section*{Data and materials availability}
The deidentified replication package accompanying this preprint contains the candidate-program manifest, inclusion decisions, frozen codebook, source capture manifest, both coding sheets, analysis code and tests, derived tables, agreement statistics, and deidentified Study 2 evidence descriptions. Unredacted institutional artifacts are withheld to reduce site-identification risk and prevent disclosure of operational identifiers.

\section*{References}
\begin{enumerate}
\small\setlength{\itemsep}{3pt}
\item Anthropic. \emph{Claude Team plan for scientists}. 2026. \url{https://claude.com/programs/team-plan-for-scientists}
\item Anthropic. \emph{Expanding our support for scientists}. 27 August 2026. \url{https://www.anthropic.com/news/expanding-support-for-scientists}
\item OpenAI. \emph{Researcher Access Program}. Captured 19 September 2026. \url{https://grants.openai.com/prog/openai_researcher_access_program/}
\item OpenAI. \emph{ChatGPT for academic researchers}. Captured 19 September 2026. \url{https://openai.com/index/chatgpt-for-academic-researchers/}
\item Anthropic. \emph{External Researcher Access Program}. Captured 19 September 2026. \url{https://support.claude.com/en/articles/9125743-what-is-the-external-researcher-access-program}
\item Anthropic. \emph{Life Sciences Verification Program}. Captured 19 September 2026. \url{https://www.anthropic.com/news/life-sciences-verification-program}
\item National Science Foundation. \emph{National Artificial Intelligence Research Resource Pilot}. Captured 19 September 2026. \url{https://www.nsf.gov/geo/updates/national-artificial-intelligence-research-resource-nairr}
\item University of Toronto Data Sciences Institute. \emph{Claude API Credit Award}. Captured 19 September 2026. \url{https://datasciences.utoronto.ca/claude-api-credit/}
\item Connecticut AI Alliance. \emph{AI Computing Access Program}. Captured 19 September 2026. \url{https://www.theccic.org/collaborations/caia/hpc/}
\item Stanford Institute for Human-Centered Artificial Intelligence. \emph{Google Cloud Credit Grants}. Captured 19 September 2026. \url{https://hai.stanford.edu/research/grant-programs/google-cloud-credit-grants}
\item Pandit HJ, Blankvoort DAH, Shaaban A, Luccioni S, Birhane A. Terms of (Ab)Use: An analysis of GenAI services. arXiv:2603.18964. 2026. \url{https://arxiv.org/abs/2603.18964}
\item Treude C. Accountable agents in software engineering: An analysis of terms of service and a research roadmap. arXiv:2605.04532. 2026. doi:10.1145/3805760.3814889
\item Bateyko D, Levy K. One Bad NOFO? AI governance in federal grantmaking. \emph{Proceedings of the 2025 ACM Conference on Fairness, Accountability, and Transparency}. 2025. doi:10.1145/3715275.3732109
\item Wang H, Dang A, Wu Z, Mac S. Generative AI in higher education: Seeing ChatGPT through universities' policies, resources, and guidelines. \emph{Computers and Education: Artificial Intelligence}. 2024;7:100326. doi:10.1016/j.caeai.2024.100326
\item Manikonda L, Outlaw D. Policy fragmentation or institutional alignment? Institutional governance of AI in universities and business schools. arXiv:2608.03584. 2026. \url{https://arxiv.org/abs/2608.03584}
\item Robert J, Muscanell N, McCormack M, Pelletier K, Arnold K, Arbino N, Young K, Reeves J. \emph{2025 EDUCAUSE Horizon Report: Teaching and Learning Edition}. EDUCAUSE; 2025.
\item AllahRakha N. Accuracy, reliability, and regulatory concerns in the use of AI tools in education and research. \emph{Social Sciences \& Humanities Open}. 2026;14:103102. doi:10.1016/j.ssaho.2026.103102
\item r/Professors. \emph{University policy on professor AI use?} 2026. \url{https://www.reddit.com/r/Professors/comments/1wpy6om/university_policy_on_professor_ai_use/}
\item Google for Education Community. \emph{How is your institution holding the AI policy and guidance?} 2026. \url{https://www.googleforeducommunity.com/t5/Discussions/How-is-your-institution-holding-the-AI-policy-amp-guidance/m-p/245173/highlight/true}
\item University of Maryland Engineering IT. \emph{Pathways to Purchase for Claude at UMD}. 2026. \url{https://ask.eng.umd.edu/162409}
\item University of Maryland Engineering IT. \emph{Claude AI}. 2026. \url{https://ask.eng.umd.edu/160855}
\end{enumerate}
\clearpage
\clearpage\section*{Supplementary information}
\subsection*{S1. Corpus disposition}
\par\noindent\begin{minipage}{\linewidth}
{\footnotesize\setlength{\tabcolsep}{4pt}\renewcommand{\arraystretch}{1.18}
\begin{tabular}{>{\raggedright\arraybackslash}p{\dimexpr 0.06462\linewidth-2\tabcolsep\relax}>{\raggedright\arraybackslash}p{\dimexpr 0.46154\linewidth-2\tabcolsep\relax}>{\raggedright\arraybackslash}p{\dimexpr 0.23538\linewidth-2\tabcolsep\relax}>{\raggedright\arraybackslash}p{\dimexpr 0.23846\linewidth-2\tabcolsep\relax}}
\toprule
\textbf{ID} & \textbf{Program} & \textbf{Included} & \textbf{Reason if excluded} \\
\midrule
P01 & OpenAI Researcher Access Program & Yes & - \\
P02 & ChatGPT for Academic Researchers & Yes & - \\
P03 & Anthropic External Researcher Access Program & Yes & - \\
P04 & Anthropic Claude Team Plan for Scientists & Yes & - \\
P05 & Anthropic AI for Science project credits & Yes & - \\
P06 & Anthropic Life Sciences Verification Program & Yes & - \\
P07 & Google Cloud Research Credits & No & Captured page did not explicitly name AI/ML, frontier models, or AI-capable compute \\
P08 & AWS Cloud Credits for Research & No & Captured page did not explicitly name AI/ML, frontier models, or AI-capable compute \\
P09 & NAIRR Pilot research resource allocations & Yes & - \\
P10 & University of Toronto DSI Claude API Credit Award & Yes & - \\
P11 & Anthropic rare-disease research grants & No & Deadline passed before capture date \\
P12 & Claude Science AI for Science projects & No & Deadline passed before capture date \\
P13 & AIRR Hartree Centre Cloud Pilot & No & Deadline passed before capture date \\
P14 & Connecticut AI Alliance AI Computing Access Program & Yes & - \\
P15 & Stanford HAI Google Cloud Credit Grants & Yes & - \\
\bottomrule
\end{tabular}
}\end{minipage}\par\vspace{6pt}
\subsection*{S2. Agreement statistics}
\par\noindent\begin{minipage}{\linewidth}
{\footnotesize\setlength{\tabcolsep}{4pt}\renewcommand{\arraystretch}{1.18}
\begin{tabular}{>{\raggedright\arraybackslash}p{\dimexpr 0.47692\linewidth-2\tabcolsep\relax}>{\raggedright\arraybackslash}p{\dimexpr 0.2\linewidth-2\tabcolsep\relax}>{\raggedright\arraybackslash}p{\dimexpr 0.32308\linewidth-2\tabcolsep\relax}}
\toprule
\textbf{Dimension} & \textbf{Exact agreement} & \textbf{Cohen's $\kappa$} \\
\midrule
D1 Unit of grant & 10/10 & 1.000 \\
D2 Stated eligibility count & 3/10 & 0.231 \\
D3 Institution-type filter & 10/10 & 1.000 \\
D4 Implied prerequisite count & 8/10 & 0.412 \\
D5 Liability assignment & 6/10 & 0.355 \\
D6 Documentation burden & 7/10 & 0.559 \\
D7 Allocation-to-use mismatch & 5/10 & 0.275 \\
D8 Term duration & 10/10 & 1.000 \\
D9 Reported success metric & 7/10 & 0.492 \\
D10 Discoverability & 10/10 & Undefined (no between-program variance) \\
\bottomrule
\end{tabular}
}\end{minipage}\par\vspace{6pt}
\subsection*{S3. Core coder-specific classifications}
\par\noindent\begin{minipage}{\linewidth}
{\footnotesize\setlength{\tabcolsep}{4pt}\renewcommand{\arraystretch}{1.18}
\begin{tabular}{>{\raggedright\arraybackslash}p{\dimexpr 0.05846\linewidth-2\tabcolsep\relax}>{\raggedright\arraybackslash}p{\dimexpr 0.10154\linewidth-2\tabcolsep\relax}>{\raggedright\arraybackslash}p{\dimexpr 0.10154\linewidth-2\tabcolsep\relax}>{\raggedright\arraybackslash}p{\dimexpr 0.1\linewidth-2\tabcolsep\relax}>{\raggedright\arraybackslash}p{\dimexpr 0.1\linewidth-2\tabcolsep\relax}>{\raggedright\arraybackslash}p{\dimexpr 0.19692\linewidth-2\tabcolsep\relax}>{\raggedright\arraybackslash}p{\dimexpr 0.17077\linewidth-2\tabcolsep\relax}>{\raggedright\arraybackslash}p{\dimexpr 0.17077\linewidth-2\tabcolsep\relax}}
\toprule
\textbf{ID} & \textbf{D5 pass 1} & \textbf{D5 pass 2} & \textbf{D7 pass 1} & \textbf{D7 pass 2} & \textbf{D8 pass 1} & \textbf{D9 pass 1} & \textbf{D9 pass 2} \\
\midrule
P01 & Not stated & Not stated & None & Unclear & 365 days & None stated & None stated \\
P02 & Not stated & Not stated & Possible & Possible & Not stated & Use & Mixed \\
P03 & Other & Shared & Unclear & Possible & Not stated & None stated & None stated \\
P04 & PI/awardee & Not stated & Yes & Possible & 365 days & None stated & None stated \\
P05 & Not stated & Not stated & Unclear & Unclear & Not stated & None stated & None stated \\
P06 & Shared & Shared & Unclear & Possible & 365 days standard; \textasciitilde{}182 days high-risk & None stated & Allocation only \\
P09 & Not stated & Not stated & Unclear & Unclear & Not stated & None stated & Allocation only \\
P10 & PI/awardee & Not stated & None & None & 365 days & Mixed & Mixed \\
P14 & PI/awardee & Shared & Unclear & Possible & Not stated & Mixed & Mixed \\
P15 & Not stated & Not stated & Unclear & Unclear & Not stated & None stated & None stated \\
\bottomrule
\end{tabular}
}\end{minipage}\par\vspace{6pt}
For the manuscript's ``no use or outcome metric'' result, an allocation-only metric is counted as no downstream use or outcome metric. No coder value is silently overwritten.

P02's nonzero D9 classifications use general published indicators of ChatGPT use in science, not outcomes measured specifically among recipients of the academic-researcher access program.

\subsection*{S4. Study 2 evidence register}
\par\noindent\begin{minipage}{\linewidth}
{\footnotesize\setlength{\tabcolsep}{4pt}\renewcommand{\arraystretch}{1.18}
\begin{tabular}{>{\raggedright\arraybackslash}p{\dimexpr 0.20769\linewidth-2\tabcolsep\relax}>{\raggedright\arraybackslash}p{\dimexpr 0.34615\linewidth-2\tabcolsep\relax}>{\raggedright\arraybackslash}p{\dimexpr 0.44615\linewidth-2\tabcolsep\relax}}
\toprule
\textbf{Evidence ID} & \textbf{Artifact} & \textbf{Public reporting rule} \\
\midrule
S2-PROGRAM-01 & Provider's public scientific team-plan terms & Cite public program page \\
S2-POLICY-01 & Faculty/staff AI acceptable-use policy, version 2.0 & Paraphrase without site identifiers; retain unredacted copy privately \\
S2-REQUEST-01 & Researcher's institutional request & Treat submission date as author-reported; report scope and question; omit institution and sender identifiers \\
S2-TICKET-01 & Automated incident receipt & Report creation of ticket; redact institution, incident number, message reference, and phone number \\
S2-CUTOFF-01 & Evidence cutoff at Day 18 & Report as right-censored if unresolved \\
\bottomrule
\end{tabular}
}\end{minipage}\par\vspace{6pt}
The policy instructed faculty and staff to contact information technology services before acquiring AI products, including free tools, and stated that vendor management would route the request for validation. The request followed that instruction and restricted proposed use to public or non-restricted scientific information. The automated receipt confirmed ticket creation. No substantive response or alternative pathway was observed by the cutoff.

\subsection*{S5. Study 2 chronology}
\par\noindent\begin{minipage}{\linewidth}
{\footnotesize\setlength{\tabcolsep}{4pt}\renewcommand{\arraystretch}{1.18}
\begin{tabular}{>{\raggedright\arraybackslash}p{\dimexpr 0.20769\linewidth-2\tabcolsep\relax}>{\raggedright\arraybackslash}p{\dimexpr 0.34615\linewidth-2\tabcolsep\relax}>{\raggedright\arraybackslash}p{\dimexpr 0.44615\linewidth-2\tabcolsep\relax}}
\toprule
\textbf{Date} & \textbf{Event} & \textbf{Observable outcome} \\
\midrule
Before Day 0 & Scientific team-plan access awarded & Workspace available at provider level \\
Day 0 (author-reported request date) & Institutional authorization request submitted & Automated incident ticket created \\
Day 18 & Evidence cutoff & No substantive response, documentation request, authorization decision, or alternative path received \\
\bottomrule
\end{tabular}
}\end{minipage}\par\vspace{6pt}
The award date is not used to compute a clearance interval because it was not verified in the dated artifacts assembled for this version. The only reported interval is 18 days from the author-reported institutional-request date to the evidence cutoff. The retained screenshots establish the request content and ticket creation but do not visibly display the submission date.

At the time of publication, the university had not replied to the request. This author-reported status update is separate from the dated 18-day process-trace interval.

\subsection*{S6. Community-context search}
Searches of indexed Reddit and community sources identified discussions of unclear university policy for professors' own AI use, fragmented institutional guidance, and an operational problem inviting researchers across different email subdomains. These observations motivated comparison but were not added to the program corpus or used to estimate frequency.

\subsection*{S7. Analysis deviations}
\begin{itemize}
\item A proposed subtraction of D4 implied prerequisites from D2 stated eligibility conditions was dropped because the counts describe different objects.
\item A proposed capability-tier analysis was not performed because no tier variable or prespecified tier-construction rule existed in the frozen inputs.
\item No pooled consensus classification was manufactured for dimensions with modest agreement. Coder-specific ranges are reported.
\end{itemize}
\subsection*{S8. Reproducibility contents}
The accompanying ancillary package contains:

\begin{itemize}
\item candidate-program and inclusion manifest;
\item frozen and final codebooks;
\item both independent coding sheets;
\item analysis script and validation tests;
\item derived corpus, agreement, and descriptive-analysis tables;
\item capture manifest and source-evidence ledger; and
\item deidentified Study 2 evidence descriptions.
\end{itemize}
Raw institutional screenshots and the unredacted institutional policy copy are not included in the public ancillary package.

\end{document}